\documentclass[aps,pra,preprint,amsmath,amssymb]{revtex4}
\usepackage{graphicx,epsfig,epsf}
\usepackage{dcolumn}
\usepackage{bm}

\begin{document}
\newcommand{\ri}{{\rm i}}
\newcommand{\re}{{\rm e}}
\newcommand{\bx}{{\bf x}}
\newcommand{\bb}{{\bf b}}
\newcommand{\bd}{{\bf d}}
\newcommand{\be}{{\bf e}}
\newcommand{\br}{{\bf r}}
\newcommand{\bk}{{\bf k}}
\newcommand{\bB}{{\bf B}}
\newcommand{\bE}{{\bf E}}
\newcommand{\bI}{{\bf I}}
\newcommand{\bJ}{{\bf J}}
\newcommand{\bR}{{\bf R}}
\newcommand{\bS}{{\bf S}}
\newcommand{\cL}{{\cal L}}
\def\Jp#1{J_+^{(#1)}}
\def\Jm#1{J_-^{(#1)}}
\newcommand{\bZero}{{\bf 0}}
\newcommand{\bM}{{\bf M}}
\newcommand{\bn}{{\bf n}}
\newcommand{\bs}{{\bf s}}
\newcommand{\tbs}{\tilde{\bf s}}
\newcommand{\rSi}{{\rm Si}}
\newcommand{\beps}{\mbox{\boldmath{$\epsilon$}}}
\newcommand{\bmu}{\mbox{\boldmath{$\mu$}}}
\newcommand{\rg}{{\rm g}}
\newcommand{\tr}{{\rm tr}}
\newcommand{\xmax}{x_{\rm max}}
\newcommand{\ra}{{\rm a}}
\newcommand{\rx}{{\rm x}}
\newcommand{\rs}{{\rm s}}
\newcommand{\rP}{{\rm P}}
\newcommand{\up}{\uparrow}
\newcommand{\down}{\downarrow}
\newcommand{\hc}{H_{\rm cond}}
\newcommand{\kb}{k_{\rm B}}
\newcommand{\cI}{{\cal I}}
\newcommand{\tit}{\tilde{t}}
\newcommand{\cE}{{\cal E}}
\newcommand{\dgtwo}{\langle \delta g^2\rangle}
\newcommand{\dgthree}{\langle \delta g^3\rangle}
\newcommand{\cC}{{\cal C}}
\newcommand{\Ubs}{U_{\rm BS}}
\newcommand{\qq}{{\bf ???}}
\newcommand*{\etal}{\textit{et al.}}
\def\vec#1{\mathbf{#1}}
\def\ket#1{|#1\rangle}
\def\bra#1{\langle#1|}
\def\keps{\mathbf{k}\boldsymbol{\varepsilon}}
\def\dm{\boldsymbol{\wp}}
\def\CG#1#2#3#4#5#6{C{\small \begin{array}{ccc}{#1}&{#3}&{#5}\\{#2}&{#4}&{#6}\end{array}}}
\def\cLL#1#2#3#4#5#6#7#8{L{\small \begin{array}{cccc}{#1}&{#3}&{#5}&{#7}\\{#2}&{#4}&{#6}&#8\end{array}}}
\def\CGprim#1#2#3#4#5#6{C^{{#1}\,{#3}\,{#5}}_{{#2}\,{#4}\,{#6}}}
\def\CLLprim#1#2#3#4#5#6#7#8{{\cal L}_1^{{#1}\,{#3}\,{#5}\,{#7}}_{{#2}\,{#4}\,{#6}\,{#8}}}

\sloppy
\title{Sub-shot noise sensitivities without entanglement}

\author{F. Benatti $^{(1,2)}$, D. Braun $^{(3,4)}$}
\affiliation{\small
{$^{(1)}$ Dipartimento di Fisica, Universit\`a di Trieste, I-34151 Trieste, Italy}\\
{$^{(2)}$ Istituto Nazionale di Fisica Nucleare, Sezione di Trieste,
I-34151 Trieste, Italy}\\
{$^{(3)}$ Laboratoire de Physique Th\'eorique  -- IRSAMC,
  Universit\'e de Toulouse, UPS, F-31062 Toulouse, France}\\
{$^{(4)}$ LPT -- IRSAMC, CNRS, F-31062 Toulouse, France}}

\centerline{\today}
\begin{abstract}
It is commonly maintained that entanglement is necessary to beat the shot noise
limit in the sensitivity with which certain parameters can be measured
in interferometric experiments.
Here we show that, with a fluctuating number of
two-mode bosons, the shot-noise limit can be beaten by
non-entangled bosonic states with all bosons in one mode. For a given finite
maximum number of bosons, we calculate the optimal one- and two-mode bosonic
states, and show that in the absence of losses, NOON states are the optimal
two-mode bosonic states.
\end{abstract}
\maketitle

\section{Introduction}

Suppose a parameter-dependent probability distribution $\mu_\theta(\xi)$
arises from the description of a classical system consisting of $N$
independent parties.
No matter what estimator one uses for estimating the parameter
$\theta$ from measured values $\xi_i$ drawn from the probability distribution,
a universal lower bound to the best mean square error in the
determination of $\theta$ is given by the inverse of the classical
Fisher information $F[\mu,\theta]$: this at best behaves
as $1/N$, a scaling known as shot-noise limit or standard quantum
limit \cite{Cramer46,Rao45}.

The field of quantum metrology concerns the use of quantum mechanical features
to improve on the above classical limitation
\cite{caves_measurement_1980,caves_quantum-mechanical_1981}. In
particular, using systems consisting of $N$ subsystems prepared in
entangled states, one can prove
that the shot-noise limit can be beaten
\cite{giovannetti_quantum_2006}. The squared sensitivity of the
determination of a parameter
$\theta$ has been proved to be bounded from below by the
inverse of a quantity known as quantum Fisher information
\cite{helstrom_quantum_1969,braunstein_statistical_1994}. The quantum
Fisher information
can scale as fast as $N^2$ if the state $\rho_\theta$ of the system
represents certain specific
$N$-partite entangled states, a scaling known as the Heisenberg limit.
Based on this, in the literature one often finds stated that, albeit
not sufficient, entanglement is necessary for overcoming the
shot-noise limit (see \cite{giovannetti_advances_2011} for a recent
review).

However, in experimental contexts where identical particles are used
for metrological
purposes, as for instance ultracold atoms trapped in double-well
potentials which can be
effectively described as two-mode bosons \cite{Oberthaler1,Treutlein},
the very notion of entanglement, that is of quantum non-locality, has
to be generalized with respect to the case when the constituent
parties, say qubits, are distinguishable.
Indeed, in this latter case there is a predetermined tensor product structure
related to the particle aspect of first quantization: for instance, for two qubits
the algebra of observables is $M_2\otimes M_2$, where $M_2$ is the
algebra of $2\times2$ matrices
for the first and second qubit, respectively.
Instead, in the case of identical particles, such a structure is no more available and
one is forced to speak of entanglement always in relation to a given algebraic context specified by a suitable mode description typical of the second quantization formalism \cite{Benatti1,Benatti2}.

In the following, we show that in the case when the number of identical bosons is
not fixed, the shot-noise limit $1/\overline{N}$ represented by
the inverse of the average boson number $\overline{N}$, can be
beaten by non-entangled states with all bosons in one mode, without
the need of partially populating the other mode. We also identify the
optimal one-mode and two mode pure states in the sense of maximum
quantum Fisher information for given maximum number of bosons and show that for two modes these are NOON
states.

\section{Quantum Metrology with Identical Particles}
\subsection{Basic quantum parameter estimation theory}
Consider a (possibly mixed) quantum state $\rho_\theta$ that depends
on the parameter $\theta$ whose value we want to find out as precisely
as possible.  $N$ repeated generalized measurements  with POVM elements
(non-negative Hermitian operators $E(\xi)$, $\int d\xi E(\xi)={\bf
  1}$) in the identically prepared state $\rho_\theta$ leads to $N$
measurement outcomes $\xi_i$ ($i=1,\ldots,N)$, distributed according
to $\mu_\theta(\xi)=\tr \rho_\theta E(\xi)$.  One estimates the value
of $\theta$ based on these $N$ outcomes $\xi_i$ with an estimator
function $\theta_{\rm est}(\xi_1,\ldots,\xi_N)$. The squared sensitivity with
which $\theta$ can be estimated from the data is defined as $\langle
(\delta \theta)^2\rangle$, where
\begin{equation}
\delta\theta=\frac{\theta_{\rm est}}{\frac{d\langle\theta_{\rm
      est}\rangle}{d\theta}}-\theta\,,
\end{equation}
and the average $\langle\ldots\rangle$ is over $\mu_\theta(\xi)$,
$\langle\theta_{\rm est}(\xi_1,\ldots,\xi_N)\rangle=\int
\left(\prod_{i=1}^N\mu_\theta(\xi_i)d\xi_i\right)\theta_{\rm
  est}(\xi_1,\ldots,\xi_N)$.

For an unbiased estimator ($\langle \theta_{\rm est}\rangle=\theta$
locally at the value of $\theta$ we are interested in, which we take
without restriction of generality
as $\theta=0$ in the following), a universal lower bound of
$\langle(\delta \theta)^2\rangle$ is provided
by the inverse of the quantum Fisher information $F[\rho]$,
\begin{equation}
\label{CR}
\langle(\delta\theta)^2\rangle \geq\  \frac{1}{F[\rho]}\ .
\end{equation}
where $F[\rho]={\rm Tr}\Big(\rho\,L^2\Big)$
and
\begin{equation}
\partial_\theta\rho_\theta\Big|_{\theta=0} = \frac{1}{2}\Big(\rho\,L\,+\,L\,\rho\Big)
\end{equation}
defines the symmetric logarithmic derivative $L$ of the
quantum state.
This so-called quantum Cram\'er-Rao bound
\cite{helstrom_quantum_1969,braunstein_statistical_1994} limits the best
sensitivity achievable for a given parameter dependent state
$\rho_\theta$, regardless of the choice of measurements and the
data-analysis, as it is optimized over all POVM measurements, in
addition to the optimization  over
all possible estimators used for the derivation of the classical
Cram\'er-Rao bound \cite{Cramer46}.  According to Fisher's theorem,
the bound can be
saturated in  the limit of $N\to\infty$ \cite{Fisher25}.
Beating the shot-noise limit, that is making
\begin{equation}
\label{shot-noise}
\langle(\delta\theta)^2\rangle<1/N
\end{equation}
necessarily requires
$F[\rho]\,>\,N$. Note that instead of measuring the same system $N$ times
with an identical initial preparation for each measurement, one can
equivalently measure once a composite system
consisting of $N$ identical subsystems in an initial product state.

The quantum Fisher information can be written as
$F[\rho_\theta]=4d_{\rm Bures}^2(\rho_\theta,\rho_{\theta+d\theta})$ in
terms of the Bures distance $d_{\rm
  Bures}(\rho,\sigma)=\sqrt{2(1-f(\rho,\sigma))}$, where the
fidelity $f(\rho,\sigma)=\tr((\rho^{1/2}\sigma\rho^{1/2})^{1/2})$
\cite{braunstein_statistical_1994,Bengtsson06}. For pure states $\rho=|\psi\rangle\langle\psi|$,
$\sigma=|\phi\rangle\langle\phi|$ the fidelity $f$ reduces to the
overlap $f(\rho,\sigma)=|\langle\psi|\phi\rangle|$.
Therefore, one has the intuitive
and information-theoretically plausible interpretation that the
distinguishability of two neighboring states whose parameters $\theta$
differ by an
infinitesimal amount $d\theta$ determines the best sensitivity with which
$\theta$ can be obtained through measurement of whatever observables.

If $\rho=\rho_{\theta=0}$ and $\rho_{d\theta}$ is created from $\rho$ through a unitary rotation with
self-adjoint generator $J=J^\dag$ from $\rho$,
\begin{equation} \label{J}
\rho\mapsto\rho_{d\theta}={\rm e}^{-i\,d\theta\, J}\,\rho_0\,{\rm e}^{i\,d\theta\, J}\,,
\end{equation}
one shows that
\begin{equation}
\label{F1}
F[\rho]=4\Delta^2_\rho
J=4(\tr\rho_0J^2-(\tr\rho_0J)^2)\equiv F[\rho,J]\ ,
\end{equation}
where in the last
step and from now on we make the dependence on the generator $J$
explicit \cite{braunstein_statistical_1994}.
For a pure, fully separable state $\rho=|\psi\rangle\langle\psi|$ of $N$ distinguishable subsystems and a
generator $J$ that is a sum of operators of the individual subsystems,
it turns out
that $\Delta^2_{\Psi}\,J\leq N/4$ \cite{braunstein_statistical_1994}. It follows that
in such a situation
entanglement is necessary to achieve sensitivities beyond the shot-noise
limit.\\

 Several ways are known by now how this limitation can be
surpassed.  The first one is the use of $k$-body interactions (also
known as non-linear schemes) which can offer a scaling
$\Delta^2_{\Psi}\,J\propto N^{2k-1}$ without entanglement (and $
N^{2k}$ with entanglement
\cite{luis_nonlinear_2004,beltran_breaking_2005,roy_exponentially_2006,luis_quantum_2007,rey_quantum-limited_2007,choi_bose-einstein_2008,napolitano_interaction-based_2011,boixo_generalized_2007}).
This requires, however, having $N$
particles all interact with each other. In some cases, such as
light induced interactions in a
Bose-condensate, interactions can be relatively naturally
induced. They typically
lead to squeezed states and resemble in this respect the earliest examples of
quantum-enhanced measurements that proposed the use of squeezed light
\cite{caves_measurement_1980,caves_quantum-mechanical_1981}.
Another way is having $N$
distinguishable subsystems
interact with a single $N+1$st system  and read
out the latter
\cite{Braun11,Braun09}.
This method has
the advantage that the system needs to accommodate only $N$ interaction
terms. Furthermore, the scaling with $N$ is stable under
local decoherence, and even decoherence itself can be used as a
signal, if the $N+1$st system is an environment.

In the following we explore a third option, namely the use of indistinguishable
particles.
Before addressing this possibility, it is necessary to stress that,
for identical particles, the notion of separability (entanglement)
cannot be given independently of the modes that are selected for the
description of the system.

\subsection{Separability and entanglement for identical particles}
Identical bosons are best addressed
within the second quantization formalism by means of the Fock
representation: we shall denote by $\vert vac\rangle$  the
vacuum state and by $a_i$, $a^\dag_i$ the annihilation and creation
operators relative to an orthonormal basis $\{\vert i\rangle\}_{i\in I}$
in the single particle Hilbert space. They satisfy the commutation
relations
$[a_i\,,\,a^\dag_j]=\delta_{ij}$,
$[a_i\,,\,a_j]=[a^\dag_i\,,\,a^\dag_j]=0$; furthermore, states $\vert n_1,
n_2, \ldots , n_k\rangle$  with $n_i$ bosons in the single particle states
$\vert i\rangle$, $i = 1, 2,\ldots,k$, are generated by acting on the
vacuum as follows
\begin{equation}
\label{Fockstate0}
\vert n_1,n_2,\ldots, n_k\rangle=
\frac{\prod_{i=1}^k (a^\dag_i)^{n_i}}{\sqrt{\prod_{i=1}^kn_i!}}\,\vert
vac\rangle\ .
\end{equation}
In quantum optics one deals with exactly the same type of multi-mode
Fock-states defined in eq.(\ref{Fockstate0}), even though their physical
meaning is somewhat different.  The quantization of the electro-magnetic
field starts with the identification of a set of orthonormal mode functions
which are solutions of the classical Maxwell-equations with appropriate
boundary conditions, such as plane waves with a given wave-vector and
polarization in the case of vacuum and periodic boundary conditions. Each of
these
modes corresponds to a harmonic oscillator due to the fact that the energy of the
electro-magnetic field is quadratic in both the electric and magnetic
fields. The multi-mode Fock state (\ref{Fockstate0}) therefore has the
meaning of a product state of a set of physical harmonic oscillators.
The index $i$  labels the harmonic oscillator (alias mode of the classical
electro-magnetic field), and arises from simple first quantization.  A state
$(a^\dag_i)^n|vac\rangle$ means the $i$-th oscillator being excited in
the $n$-th one of its excited states, i.e. it~corresponds to $n$ photons in mode $i$ (see
e.g.~\cite{Scully97} or any other text-book on quantum optics).
This is in contrast to the second quantization formalism for massive identical
bosons, where $i$ labels different single particle orthonormal basis vectors; in this case, $(a^\dag_i)^n|vac\rangle$ means that
$n$ identical bosons are created in the $i$-th single particle state.
In the Bose-Hubbard approximation, instances of single atom orthonormal bases for
ultracold atoms trapped by a double-well potential are states corresponding to an atom being
localized in either one or the other of the two wells or,
the first two energy eigen-states of the single-particle Hamiltonian.
Despite the different
physical meaning, the
formalism is exactly the same in both cases, and we will therefore not
distinguish between massive identical bosons treated in second quantization
and photons in quantum optics, but have both situations in mind when we
speak of ``indistinghuishable particles''.

In the following we shall be dealing with identical bosons that can be found
in two modes identified by pairs of creation and annihilation operators
$a,a^\dag$, respectively $b,b^\dag$ satisfying the canonical commutation
relations
$[a,a^\dag]=[b,b^\dag]=1$, while the remaining ones all vanish. We shall
consider the Fock representation based on a vacuum state $\vert vac\rangle$
so that the states 
\begin{equation}
\label{Fockstate2}
\vert n_a,n_b\rangle=\frac{(a^\dag)^{n_a}(b^\dag)^{n_b}}{\sqrt{n_a!n_b!}}\,\vert vac\rangle\quad n_{a,b}\in\mathbb{N}
\end{equation}
constitute the orthonormal basis of eigenstates of the Fock number operator
$a^\dag a+b^\dag b$  with $n_a$ bosons in one mode and $n_b$ bosons in
the other one.

In the second quantization formalism there is no pre-defined algebraic
tensor product structure as for distinguishable particles; in the latter
case, one starts out with the tensor product of the algebras of operators
acting on the Hilbert spaces of the single particles: for instance, in the
case of one qubit the operator algebra is the $2\times 2$ complex matrix
algebra $M_2$ and in the case of two distinguishable qubits, it is the
$4\times 4$ matrix algebra $M_2\otimes M_2$.

In absence of a definite tensor product structure, a new approach to
locality (of observables) and separability (of states) must be
developed~\cite{Benatti1,Benatti2}: observe that the main property of local
observables $A\otimes 1$ and $1\otimes B$ for a bi-partite system consisting
of distinguishable particles is that they commute. In the case of two
bosonic modes, the tensor product structure can thus be replaced by pairs of
commuting sub-algebras $(\mathcal{A},\mathcal{B})$ generated by
$\{a,a^\dag\}$, respectively $\{b,b^\dag\}$ whereby operators of the form
$A\,B$ with $A\in\mathcal{A}$ and $B\in\mathcal{B}$ can be termed local
with respect to the pair $(\mathcal{A},\mathcal{B})$, or
$(\mathcal{A},\mathcal{B})$-local.
Furthermore, one can extend the notion of separability as follows: states
$\omega$ on the Bose algebra $\mathcal{M}$ of the two-mode system are
generic expectations (linear positive and normalized functionals)
$\mathcal{M}\ni X\mapsto\omega(X)\in\mathbb{C}$. Then, a state $\omega$ will
be called separable with respect to the pair
$(\mathcal{A},\mathcal{B})$, or $(\mathcal{A},\mathcal{B})$-separable, if,
on local observables, it splits into a convex combinations of products of
expectations with respect to other states, namely if
$$
\omega(A\,B)=\sum_i\lambda_i\,\omega_i^{(1)}(A)\,\omega^{(2)}_i(B)\ ,
$$
for all $A\in\mathcal{A}$ and $B\in\mathcal{B}$.

The simplest examples of $(\mathcal{A},\mathcal{B})$-separable states are
the Fock states in~(\ref{Fockstate2}); indeed, expectations of
$(\mathcal{A},\mathcal{B})$-local observables $A\,B$, $A\in\mathcal{A}$,
$B\in\mathcal{B}$ factorize, 
$$
\langle n_a,n_b \vert\,A\,B\,\vert n_a,n_b\rangle= \langle n_a,n_b
\vert\,A\,\vert n_a,n_b\rangle\,
\langle n_a,n_b \vert\,B\,\vert n_a,n_b\rangle\ ,
$$
and show no correlation among these commuting observables.

The definitions of locality and separability given above reduce to the standard ones in the case of two distinguishable particles; in this case one identifies the sub-algebras $\mathcal{A}$ and $\mathcal{B}$ with the single particle algebras of operators, $\mathcal{M}$ with their tensor product and the states $\omega(X)$ with the expectations
${\rm Tr}(\rho\,X)$ corresponding to density matrices $\rho$.

What should be remarked is that, while in the case of distinguishable particles the tensor product structure is somewhat taken for granted and one need not specify that locality and separability always refer to it, it is not so in the case of identical bosons: in such a case it must always be specified with respect to which pair $(\mathcal{A},\mathcal{B})$ a state is separable.
Indeed, it is easy to see that Bogolubov transformations, as those
implemented by beam-splitters either in quantum optics or in cold atom
interferometry, transform the mode operators $\{a,a^\dag\}$, $\{b,b^\dag\}$
into new mode operators $\{c,c^\dag\}$, $\{d,d^\dag\}$ such that the
$(\mathcal{A},\mathcal{B})$-separable state in~(\ref{Fockstate2}) turns out
not to be separable with respect to the new pair of commuting sub-algebras
generated by $\{c,c^\dag\}$ and $\{d,d^\dag\}$.
Other definitions of
entanglement have been discussed in the literature, e.g.~in the context of
``generalized entanglement'' (see
\cite{barnum_generalization_2005,eckert_quantum_2002,amico_entanglement_2008}
and references therein).

\subsection{Fock states}
In order to illustrate the physical consequences of the mode-dependent formulation of locality and entanglement,
we show that entanglement is not necessary for beating the shot noise limit when one deals with
identical particles.
To this end, consider an $(\mathcal{A},\mathcal{B})$-separable Fock state as
in~(\ref{Fockstate2}) with $k$ bosons in mode $a$ and $N-k$ bosons in mode
$b$,
$$
\vert k, N-k\rangle =\frac{(a^\dag)^k (b^\dag)^{N-k}}{\sqrt{k!(N-k)!}}\vert vac\rangle\,.
$$
We call a state with $k=N$ or $k=0$ a one-mode
state, as all
bosons are in the first or second mode, respectively.

Let us now consider the Schwinger-representation that associates to the two-mode bosons the
angular-momentum-like operators
\begin{equation}
\label{pseudo-spin}
J_x=\frac{a^\dag b+a  b^\dag}{2}\ , \ J_y=\frac{a^\dag b-a  b^\dag}{2i}\ ,\
J_z=\frac{a^\dag a-b^\dag b}{2}\ .
\end{equation}
A pseudo-rotation $\displaystyle U(\theta)={\rm
  e}^{i\theta\,J_x}$
is $(\mathcal{A},\mathcal{B})$-non-local for $U(\theta)$ as it cannot
  be split into the product of some $A\in\mathcal{A}$ and
  $B\in\mathcal{B}$.
Indeed, the action of $U(\theta)$ on the
  $(\mathcal{A},\mathcal{B})$-separable state
  $\vert k,N-k\rangle$ makes the state
$(\mathcal{A},\mathcal{B})$-entangled.
Furthermore, unlike for distinguishable qubits, the corresponding
  Fisher information,
\begin{eqnarray}
\nonumber
F[\vert k,N-k\rangle,J_x]&=&\,4\,\Delta_{\vert k,N-k\rangle}^2\,J_x=4\langle k,N-k\vert J^2_x\vert k,N-k\rangle\\
\label{Fisher1}
&=&N\,(2\,k\,+\,1)\,-\,2\,k^2\ ,
\end{eqnarray}
exceeds $N$ for all $k\neq 0$ and $k\neq N$.

Therefore, except when all identical bosons are in one mode and none
in the other, despite the separability of the Fock state, one can beat
the shot-noise limit profiting from the non-locality of the rotation
generated by $J_x$.

In the following section we show that state-entanglement is not
necessary to beat the shot-noise limit even when, differently from the
case just discussed, all identical qubits are in one of the two modes.

\section{Phase estimation with a Mach-Zehnder interferometer}
In this section, we identify the identical bosons with photons in
a Mach-Zehnder interferometer (MZ); notice, however, that the same
formalism applies to matter
interferometry based on trapped ultracold atoms, which is also currently
pursued (see for instance~\cite{Bloch}).

Let us consider a MZ consisting of two equal beam-splitters (BS); the first BS
generates a Bogoliubov rotation of the two modes $a$ and $b$ associated with
the two arms of the interferometer,
\begin{eqnarray}
\label{bog0a}
U_{BS}(\alpha)\,a\,U_{BS}(-\alpha)&=&a\,\cos|\alpha|\,-\,b\,{\rm e}^{i\arg(\alpha)}\,\sin|\alpha|\\
\label{bog0b}
U_{BS}(\alpha)\,b\,U_{BS}(-\alpha)&=&b\,\cos|\alpha|\,-\,a\,{\rm e}^{i\arg(\alpha)}\,\sin|\alpha|\ ,
\end{eqnarray}
with $\alpha$ a complex parameter characteristic of the $BS$, via the unitary operator
\begin{equation}
\label{bog1}
U_{BS}(\alpha)={\rm e}^{\alpha\, a^\dag b-\alpha^*\,a\,b^\dag}\ .
\end{equation}
After the first BS, along the arm of the MZ described by $a,a^\dag$, a unitary rotation by an angle $\phi$ is generated by the number operator $a^\dag\, a$.
This is in turn followed by a second BS which recombine the beams by means of $U_{BS}(-\alpha)$.

The total effect on an incoming state $\rho$ is described by the unitary operator
\begin{eqnarray}
\nonumber
U_{MZ}(\alpha,\phi)&=&U_{BS}(-\alpha)\,{\rm e}^{i\phi\,a^\dag\,a}\,U_{BS}(\alpha)={\rm e}^{i\phi\,J}\\
\nonumber
J&=&
\Big(\cos^2|\alpha|\,a^\dag\,a+\sin^2|\alpha|\,b^\dag\,b
\\
\label{bog2}
&&\hskip .5cm
-\cos|\alpha|\sin|\alpha|(
{\rm e}^{i\,arg(\alpha)}a^\dag\,b-{\rm e}^{-i\,arg(\alpha)}\,a\,b^\dag)\Big)\ .
\end{eqnarray}

\subsection{One mode photon states}
Suppose a system of identical bosons is prepared as input to the interferometer in a state
of the form
\begin{equation}
\label{input}
\vert\Psi\rangle=\sum_kc_k\,\vert k,0\rangle\ ,\quad \vert k,0\rangle=\frac{(a^\dag)^k}{\sqrt{k!}}\,\vert vac\rangle\ ,
\end{equation}
with all bosons in mode $a$, that is $a^\dag\,a\vert
k,0\rangle=k\,\vert k,0\rangle$,
$b^\dag\,b\vert k,0\rangle=0$. We shall only demand that the mean boson number
\begin{equation}
\label{mpn}
\overline{N}=\langle\Psi\vert\,(a^\dag\, a+b^\dag\, b)\,\vert\Psi\rangle=\sum_k\,p_k\,k
\end{equation}
be finite, where $p_k=|c_k|^2$, $\sum_kp_k=1$.
Using~(\ref{F1}), the quantum Fisher information corresponding to such a state and the generator $J$ in~(\ref{bog2}) is readily computed:
\begin{eqnarray}
\nonumber
F[\Psi,J]&=&
=4\Big[\cos^4|\alpha|\,\Bigg(\sum_kp_k\,k^2\,-\,\Big(\sum_kp_k\,k\Big)^2\Bigg)\ +\\
\label{Fisher3}
&&\hskip 2cm
+\,\cos^2|\alpha|\,\sin^2|\alpha|\,\sum_kp_k\,k\Big]\ .
\end{eqnarray}
The term in parenthesis in the first line is the mean square error of
a stochastic variable $X$ taking values on the natural numbers
$X\in\mathbb{N}$ distributed according to the probabilities
$p(X=k)=p_k$. The variance is always non-negative; whence, choosing
$\alpha=\pi/4$ yields $F[\Psi,J]\geq \overline{N}$.
This already indicates the possibility of beating the shot-noise
limit, that is the  bound~(\ref{shot-noise}) with the mean photon
number $\overline{N}$ in the place of $N$.

Indeed, consider a balanced BS ($\alpha=\pi/4$) and choose an input
state with a finite fixed maximum number $K$ of bosons,
\begin{equation}
\label{state0}
\vert\Psi\rangle=\sum_{k=0}^K c_k \vert k,0\rangle\ .
\end{equation}
By isolating in~(\ref{Fisher3}) the term $k=K$ and using
$K\,p_K= \overline{N}- \sum_{k=1}^{K-1}p_k\,k$, one gets
$$
F[\Psi,J]=\overline{N}\Big(1+K-\overline{N}\Big)\,+\,\sum_{k=1}^{K-1}p_k\,k(k-K)\,.
$$ 
Each term in the last sum is negative.
The quantum Fisher information is thus optimized by choosing $p_k=0$
for all
$k\neq 0,K$, which in turn implies $p_K=\overline{N}/K$ and
\begin{equation}
\label{state1}
\Big\vert\Psi\Big\rangle=\sqrt{1-\frac{\overline{N}}{K}}\ \Big\vert 00\Big\rangle\,+\,{\rm e}^{i\chi}
\sqrt{\frac{\overline{N}}{K}}\ \Big\vert[K,0\Big\rangle\ .
\end{equation}
Then,
\begin{equation}
\label{F2}
F[\Psi,J]=\overline{N}\Big(K-\overline{N}+1\Big)\ .
\end{equation}
Since the Fisher
information is larger than the mean particle number, it thus follows
that the shot-noise limit can be beaten
by choosing a suitable $\overline{N}$.  A similar conclusion has been reached in~\cite{Rivas11}, where,
differently from here, the authors considered a superposition of the vacuum
state with a squeezed
state acted upon by a rotation generated by the number operator.
Earlier work on optimizing states for minimal phase uncertainty, notably in
the context of squeezed states, can be found
in
\cite{schleich_exponential,Bialynicki-Birula93,Freyberger94,ou_fundamental_1997,dariano_feasible_1996}.
Most of this earlier work used the notion of an approximate phase
operator. In \cite{Pinel11} it was shown that optimal sensitivity of a
Mach-Zehnder interferometer using
multimode Gaussian states can be reached without entanglement by
appropriate mode-engineering.

One might wonder about the importance of the scaling with the average
boson number $\overline{N}$ instead of $N$. Of course, $N$ is only
well defined for a Fock state, whereas for all other states one has to
live with fluctuating $N$. For laser light in a coherent state with
$\overline{N}\gg 1$ or even
the most squeezed states currently available \cite{Mehmet10,Keller08},
the fluctuations of $N$ still
satisfy  $\sigma(N)/N \ll 1$, and the average photon number is
therefore representative of the photon number in any
realization. A state of the form (\ref{state1}) with
$\overline{N}\simeq K/2$ maximizes the photon-number fluctuations,
however, and obviously the average value of $N$ is never realized
(only $N=0$ or $N=K$ are). Nevertheless, the scaling with
$\overline{N}$ is highly relevant practically, as it corresponds to
the mean energy in the state, which is indeed what makes producing the
state costly.  In \cite{schleich_exponential} a state was proposed
that leads to
a very sharp maximum in the distribution of measured rotation angles,
suggesting even exponential scaling of the phase uncertainty with
$\overline{N}$.

>From a physical perspective it makes sense that the optimal state
leads to maximum uncertainty in $N$, as, for a Heisenberg-uncertainty
limited state, this corresponds to minimal uncertainty in $\theta$. Since
there is no entirely satisfying definition of a phase operator,
``Heisenberg-uncertainty limited'' means here a state that satisfies the
Cram\'er-Rao bound.  Indeed, inequality (\ref{CR}) has been understood from
the beginning as a generalization of Heisenberg's uncertainty relation
\cite{braunstein_statistical_1994}. Eq.(\ref{F1}) shows that the relevant
``complementary
observable'' is the generator $J$ of eq.(\ref{J}). Furthermore, for the
one-mode states considered here, the fluctuations of $J$ are given by the
fluctuations of $N$.  Therefore the optimal one mode state must be indeed
the state that maximizes the photon number fluctuations in that mode.
The single mode ``ON'' state $(|0\rangle+|N\rangle)/\sqrt{N}$ was also
identified as the optimal state for obtaining the best possible
sensitivity of mass measurements with nano-mechanical harmonic
oscillators \cite{Braun11.2}.  It has a Wigner function with $N$ lobes in
azimuthal direction.

\subsection{Optimal state}
One might wonder what state is optimal for a given maximum photon
number. In \cite{durkin_local_2007} numerical and some analytical evidence was
shown that the so-called NOON-state \cite{boto_quantum_2000} is the optimal
state when losses are neglected. It has meanwhile become
clear that NOON states are not very useful in practice, as the
slightest chance of photon loss leads, in the limit of large $N$, back
to the standard quantum limit
\cite{huelga_improvement_1997,Kolodynski10,escher_general_2011}.  The
analysis in \cite{durkin_local_2007} was based on
the classical Fisher information and photon counting measurements at
the two output ports of the MZ.  Here we give
a simple demonstration
using the quantum Cram\'er-Rao bound that the NOON state in absence of
losses and dephasing is optimal
for a balanced MZ-interferometer
no matter what measurement is performed in the end.

As before, we use the Schwinger-representation~(\ref{pseudo-spin}). For a
balanced MZ-interferometer ($\alpha=\pi/4$) eq.~(\ref{bog2}) gives
\begin{equation}
\label{Jab}
J=(a^\dagger a+b^\dagger b+a^\dagger b+ab^\dagger)/2=\frac{\hat{N}}{2}+J_x\,,
\end{equation}
where $\hat{N}=a^\dagger a+b^\dagger b$ is the total photon number
operator. It proves convenient to introduce a new class of orthogonal states adapted to the Schwinger representation:
these are the pseudo-angular momentum states $|jm\rangle_\ell$, $\ell=x,y,z$, such that
\begin{equation}
\label{angstates}
J_\ell\vert j,m\rangle_\ell=m\,\vert j,m\rangle_\ell\ ,\quad \bJ^2\vert j,m\rangle_\ell= j(j+1)\,\vert j,m\rangle_\ell
\ .
\end{equation}
One easily shows that $\bJ^2=(\hat{N}/2)(\hat{N}/2+1)$,
which shows that the usual pseudo-angular momentum states $|jm\rangle_\ell$ are also eigenstates of $\hat{N}$,
$\hat{N}|jm\rangle_\ell=2j|jm\rangle_\ell$. Furthermore, if $\ell=z$,
using the expression of $J_z$ in eq.~(\ref{pseudo-spin}), the
Fock states $\vert k,N-k\rangle$ can be recast as common
eigenstates $|jm\rangle_z$ of $\bJ^2$ and $J_z$ with
\begin{equation}
\label{angstates1}
j=\frac{N}{2}\ ,\qquad -j\leq m=k-\frac{N}{2}\leq j\ .
\end{equation}

Consider now first a state $|\psi\rangle$ with fixed $j$. It is useful to
write $|\psi\rangle$ in the $J_x$ eigenbasis,
$|\psi\rangle=\sum_{m=-j}^jc_m|jm\rangle_x$, with
$J_x|jm\rangle_x=m|jm\rangle_x$.  There are no fluctuations from
$\hat{N}/2$ in $J$, and from eq.~(\ref{Jab}) we have thus $\Delta_\psi^2 J=\langle
J_x^2\rangle-\langle J_x\rangle^2$. Inserting the expansion of
$|\psi\rangle$ in the $J_x$ eigenbasis, we are led to
\begin{equation} \label{delj}
\Delta_\psi^2 J=\sum_{m=-j}^jp_m m^2-\left(\sum_{m=-j}^jp_m m\right)^2\,,
\end{equation}
with $p_m=|c_{m}|^2$.

Let $\Delta_\pi$ denote the right hand side of eq.~(\ref{delj}) depending on the distribution $\pi=\{p_m\}_{m=-j}^j$;
because of the apparent symmetry of the expression, the
distribution $\pi'=\{p_{-m}\}_{m=-j}^j$ obtained by exchanging $p_m$
with $p_{-m}$ leads to $\Delta_{\pi'}=\Delta_\pi$.
Let us then consider the symmetrized distribution $\displaystyle\pi_{sym}=\left\{\frac{p_m+p_{-m}}{2}\right\}_{m=-j}^j$; convexity yields
$$
\left(\sum_{m=-j}^j\frac{p_m+p_{-m}}{2} m\right)^2\leq
\frac{1}{2}\left(\sum_{m=-j}^jp_m m\right)^2+\frac{1}{2}\left(\sum_{m=-j}^jp_{-m} m\right)^2\ ,
$$
whence $\Delta_{\pi_{sym}}\geq\Delta_\pi$, i.e.~a generic distribution $\pi$ cannot provide a $\Delta_\pi$ larger than
the $\Delta_{\pi_{sym}}$ obtained by symmetrizing it.
The maximum $\Delta_\pi$ must then be attained at
distributions with the property that $p_m=p_{-m}$; this means maximizing
$\Delta_\psi^2 J=2\sum_{m=1}^jp_m m^2$ under the constraints $0\le p_m\le 1/2$ for all $m$ and
$\sum_{m=-j}^jp_m=1$. Since $\langle \psi|J^2_x|\psi\rangle\leq j^2$, for fixed $j$, the maximum of the variance is attained at $p_j=1/2=p_{-j}$.

Next, consider a general state with at most $j_{\rm max}$ excitations,
\begin{equation} \label{}
|\psi\rangle=\sum_{j=0}^{j_{\rm max}}\sum_{j=-m}^m c_{jm} |jm\rangle_x\,.
\end{equation}
Since both $\hat{N}$ and $J_x$ conserve $j$, we get
\begin{equation} \label{}
\langle J \rangle=\sum_j\sum_{m,n}c_{jn}^*c_{jm}\ {}_x\hskip-.05cm\langle jn|J|jm\rangle_x\,,
\end{equation}
and similarly for $J^2$. It is useful to introduce the notation $\langle X\rangle_j={}_j\langle\psi\vert X\vert\psi\rangle_j$, where
$|\psi\rangle_j$ is the wave-function in the $j$ sector, that is $|\psi\rangle_j=\sum_m
c_{jm}|jm\rangle_x/\sqrt{p_j}$, with $p_j=\sum_m |c_{jm}|^2$ assuring the
correct normalization. This gives
\begin{equation} \label{}
\Delta_\psi^2 J=\sum_j p_j\langle J^2\rangle_j-\left(\sum_j p_j\langle
J\rangle_j \right)^2=\sum_j p_j\langle\left( J-\langle J\rangle_j\right)^2
\rangle_j\,.
\end{equation}
Since $\max_{\{c_{jm}\}}\langle\left( J-\langle J\rangle_j\right)^2
\rangle_j=j^2$ grows monotonically with $j$, we obtain the
maximum of $\Delta_\psi^2 J$ over all $c_{jm}$ by choosing $p_{j_{\rm
    max}}=1$ (and correspondingly all other $p_j=0$). Thus, the state that
maximizes the quantum Fisher information is
\begin{equation}
\label{maxstate}
|\psi\rangle=\frac{1}{\sqrt{2}}\left(|j_{\rm max}j_{\rm
 max}\rangle_x+e^{i\chi}|j_{\rm max}\,-j_{\rm max}\rangle_x\right) \,,
\end{equation}
where $\chi$ is an arbitrary phase.

In order to give a physical interpretation to such states, let us consider the Bogolubov transformation
\begin{equation}
a=\frac{c+d}{\sqrt{2}}\ ,\quad b=\frac{c-d}{\sqrt{2}}
\end{equation}
to new modes described by creation (annihilation) operators $c\,,\, d$
($c^\dagger$, $d^\dagger$).
With reference to the new modes, the pseudo angular momentum operator $J_x$ becomes
\begin{equation}
\label{Jcd}
J_x=\frac{c^\dag c-d^\dag d}{2}\ .
\end{equation}
In terms of occupation number states of these modes, it follows that the state~(\ref{maxstate})
has the form of a NOON state (see~(\ref{angstates1}), namely a superposition of all photons in mode $c$ and all photons in mode $d$:
\begin{equation}
\label{NOONstate}
|\psi\rangle=\frac{(c^\dag)^{2j_{max}}\,+\,e^{i\chi}\,(d^\dag)^{2j_{max}}}{\sqrt2}\,\vert0\rangle\ .
\end{equation}

Besides to photons, the above analysis also applies to interferometric
setups based on ultracold atom gases trapped by  double well
potentials: the modes $a,\, b$ describe atoms confined in the left and
right well, whereas the modes $c,\,d$ are related to the first
two tunneling split single particle energy eigenstates (in the limit
of high barrier).

\section{Conclusions}

Entanglement is not necessary for beating the shot-noise limit
when using identical bosons.
In the case of fixed boson number $N$, states $\vert k, N-k\rangle$
with $k$ bosons in mode $a$ and $N-k$ in mode $b$ are separable with
respect to these modes in the sense that there are no correlations
between observables of the first and the second mode.
Nevertheless, unless when $k=0,N$ (the single mode case), these states
can achieve squared sensitivities scaling faster than $1/N$ if
subjected to beam
splitting transformations generated by pseudo angular momentum
operators like $J_x=(a^\dag\, b\,+\,a\,b^\dag)/2$ that are non-local
with respect to the given modes. Superpositions of Fock-states can
beat the shot-noise limit even for single mode states. An ``ON'' state
in one mode, i.e.~a superposition of 0 and N photons in one mode,
leads to a scaling of the squared sensitivity of a Mach-Zehnder
interferometer as $\propto
1/\overline{N}^2$, which corresponds to the Heisenberg limit.  The
latter scaling is also obtained, at least in principle under ideal
unitary evolution, for the NOON state.  Using the quantum Fisher
information we showed that the NOON state is --- in such a highly
idealized situation --- the optimal two mode state, in the sense that
it saturates the quantum Cram\'er-Rao bound for pure states with the
same
maximum number of excitations fed into a Mach-Zehnder interferometer.

\acknowledgments DB would like to thank Wolfgang Schleich for interesting
  discussions.

\bibliography{mz_bt_v2}

\end{document}